# Estimating the variance-covariance matrix of two-step estimates of latent variable models: A general simulation-based approach

Roberto Di Mari*

Department of Economics and Business

University of Catania, Italy

and

Jouni Kuha

Department of Statistics

London School of Economics and Political Science, London, UK

July 21, 2025

**Abstract**

We propose a general procedure for estimating the variance-covariance matrix of two-step estimates of structural parameters in latent variable models. The method is partially simulation-based, in that it includes drawing simulated values of the measurement parameters of the model from their sampling distribution obtained from the first step of two-step estimation, and using them to quantify part of the variability in the parameter estimates from the second step. This is asymptotically equal with the standard closed-form estimate of the variance-covariance matrix, but it avoids the need to evaluate a cross-derivative matrix which is the most inconvenient element of the standard estimate. The method can be applied to any types of latent variable models. We present it in more detail in the context of two common models where the measurement items are categorical: latent class models with categorical latent variables and latent trait models with continuous latent variables. The good performance of the proposed procedure is demonstrated with simulation studies and illustrated with two applied examples.

---
*Email: roberto.dimari@unict.it – address: Corso Italia 55, 95128, Catania, Italy.

Di Mari acknowledges funding by the European Union - NextGenerationEU, Mission 4, Component 2, in the framework of the GRINS -Growing Resilient, INclusive and Sustainable project (GRINS PE00000018 – CUP E63C22002120006). The views and opinions expressed are solely those of the author and do not necessarily reflect those of the European Union, nor can the European Union be held responsible for them.

# 1  Introduction

General latent variable models combine two basic elements: *measurement models* for how the latent variables are related to observed measures (items) of them, and *structural models* for the relationships among the latent variables and observed explanatory and response variables for them. Different types of latent variable models are obtained with different specifications for the variables and models in this general formulation. For example, in this paper we present the general method of variance estimation that we propose in the context of two families of models, in both of which the items are categorical variables but the latent variables are of different types: latent class models where they are categorical and latent trait models (item response theory models) where they are continuous variables. We suppose that the focus of interest is the structural model, as is often the case in substantive applications.

Parameters of such models can be estimated in different ways. We contrast, in particular, two broad approaches: *one-step estimation* where the parameters of both the measurement and structural models are estimated at once, and *two-step estimation* where they are estimated in separate steps. Specifically, in the first step of the two-step approach the measurement parameters are estimated from the simplest model that allows this, and in the second step the structural parameters are estimated with the measurement parameters held fixed at their estimated values from the first step. With likelihood-based estimation, which will be our focus, one-step estimates are standard maximum likelihood (ML) estimates with their usual properties, and two-step estimation is an instance of general pseudo-ML estimation in the sense defined by Gong & Samaniego (1981).

Two-step estimation has been developed for different families of latent variable models, for linear structural equation models (SEMs) by Rosseel & Loh (2024), for latent class models by Bakk & Kuha (2018), and for latent trait models by Kuha & Bakk (2023), and has been extended to further situations, e.g. to multilevel latent class models by Di Mari et al. (2023). Theoretical considerations and simulation studies show consistently that it works well, with essentially similar (or even better) finite-sample performance as one-step estimation. The two-step approach also has some further advantages. Conceptually, by separating the estimation of the measurement and structural models it also separates the effective *definition* of the latent variables from what they are used for, and thus avoids confusion ("interpretational confounding") between the two. Practically, it can be computationally much simpler than one-step estimation, for example when the same estimated measurement model from step 1 can be used with several different strucrural models in step 2.

However, standard two-step estimation has one element that mars its full convenience. This appears in variance estimation. In general, variances of two-step estimates of structural parameters should include two contributions, reflecting respectively the uncertainties in the estimated measurement and structural parameters from steps 1 and 2. The asymptotic pseudo-ML expression for their variance-covariance matrix (as shown in Section 4.1 below) is a sum of two terms accordingly. The inconvenient piece in this is a matrix of second-order derivatives of the log likelihood, with respect to a cross of the two sets of parameters. This is the only part of the two-step procedure which does not come directly from either of its two steps. Instead, it has to be calculated separately. This has been implemented in some software for two-step estimation, such as the R packages `lavaan` (Rosseel, 2012) for SEMs and `multilevLCA` (Lyrvall, Di Mari, et al., 2025) for some single-level and multilevel latent class models. For other models it may require substantial additional effort, either explicitly deriving and coding the necessary derivatives, or a further call to estimation software which tricks it into calculating this matrix (for examples in this and other contexts, see e.g. Skrondal & Kuha, 2012; Oberski & Satorra, 2013; Bakk & Kuha, 2018; Di Mari & Bakk, 2018; Di Mari et al., 2020; Di Mari et al., 2023; Kuha & Bakk, 2023).

What we propose in this paper is using parameter simulation to get around this complication. From step 1 we now take forward not only the point estimates of the measurement parameters but also a set of simulated valued for them, drawn from their estimated sampling distribution from this step. Step 2 is then carried out not only once given the point estimates of the measurement parameters (to get the point estimates of the structural parameters) but also given each of the simulated values of them in turn. Observed variation over the latter can be used to capture the contribution to the variance-covariance matrix from the uncertainty in the step-1 estimates. It can be shown that this gives variance estimates which are asymptotically equivalent to the ones from the standard matrix formulas, and our simulation studies also indicate that they perform essentially similarly. No separate steps for calculating the invonvenient cross-derivative matrix are now needed. Instead, the proposed method restores a pure two-step procedure where its two steps are fully separated.

While we focus here on likelihood-based estimation, the ideas of both two-step estimation and our simulation-based variance estimation can also be used, essentially unchanged, with other methods of estimation. This also also provides the closest (and only) example of comparable variance estimation for latent variable models in previous literature. This has been proposed by Levy (2023) and Levy & McNeish (2025), who carry our both steps of estimation and the simulations from the first step in the Bayesian framework, using MCMC estimation. Our proposal in this paper is essentially a likelihood-based version and justification of the same idea, with the simulations drawn from an estimated frequentist sampling distribution of the measurement parameters.

In the rest of the article, in Section 2 we define the general latent variable model to which these methods can be applied, and the latent class and latent trait models that we will consider in more detail. In Section 3 we introduce the basic ideas of two-step estimation. Methods of variance estimation, both the standard asymptotic formulas and our proposed simulation-based approach, are described in Section 4. Simulation studies of the performance of the methods are given in Section 5, two applied examples in Section 6, and concluding remarks in Section 7.

## 2 The model setup: A general latent variable model

Let $\boldsymbol{\eta}$ denote a vector of latent variables, and $\mathbf{Y}$ and $\mathbf{Z}$ two distinct vectors of observed variables for a unit of analysis (we omit for the moment a unit subscript from the notation). Consider a model of the form $p(\mathbf{Y}, \boldsymbol{\eta}, \mathbf{Z}; \boldsymbol{\theta}) = p(\mathbf{Y}|\boldsymbol{\eta}, \mathbf{Z}; \boldsymbol{\theta}_1) \, p(\boldsymbol{\eta}, \mathbf{Z}; \boldsymbol{\theta}_2)$, where $p(\cdot|\cdot)$ denotes a conditional distribution and $\boldsymbol{\theta} = (\boldsymbol{\theta}_1', \boldsymbol{\theta}_2')'$ are parameters, with $\boldsymbol{\theta}_1$ and $\boldsymbol{\theta}_2$ assumed to be distinct and variation-independent of each other. The model for the observed variables implied by this is

$$p(\mathbf{Y}, \mathbf{Z}; \boldsymbol{\theta}) = \int p(\mathbf{Y}|\boldsymbol{\eta}, \mathbf{Z}; \boldsymbol{\theta}_1) \, p(\boldsymbol{\eta}, \mathbf{Z}_i; \boldsymbol{\theta}_2) \, d\boldsymbol{\eta}. \tag{1}$$

We refer to $p(\mathbf{Y}|\boldsymbol{\eta}, \mathbf{Z}; \boldsymbol{\theta}_1)$ as the *measurement model* and $p(\boldsymbol{\eta}, \mathbf{Z}; \boldsymbol{\theta}_2)$ as the *structural model* of the joint model (1). The variables $\mathbf{Y}$ (the measurement *items*) are regarded as observed measures of the latent $\boldsymbol{\eta}$, while $\mathbf{Z}$ are observed explanatory and/or response variables in the structural model.

This defines a very general model. All common families of latent variable models are instances of it, obtained with different specifications for the distributions in (1). The idea of two-step estimation applies to all of these models, and the method of variance estimation that we propose here could be used for any of them. For specificity and notational simplicity, however, in the examples of this paper we focus on some specific instances of this general model. We consider models where $\mathbf{Z} = (\mathbf{Z}_p', \mathbf{Z}_o')'$, of which $\mathbf{Z}_p$ appear in the structural model only as explanatory variables (covariates) for (some or all of) $(\boldsymbol{\eta}', \mathbf{Z}_o')'$, and $\mathbf{Z}_o$ as response variables to $(\mathbf{Z}_p', \boldsymbol{\eta}')'$. Here $\mathbf{Z}_p$ and/or $\mathbf{Z}_o$ may also be empty, in which case they would be omitted from the notation. The structural model can then be considered conditionally on $\mathbf{Z}_p$, so that the marginal distribution of $\mathbf{Z}_p$ is omitted from the estimation and its parameters are not included in $\boldsymbol{\theta}_2$. The structural model of interest is then

$$p(\boldsymbol{\eta}, \mathbf{Z}_o|\mathbf{Z}_p; \boldsymbol{\theta}_2) = p(\mathbf{Z}_o|\boldsymbol{\eta}, \mathbf{Z}_p; \boldsymbol{\theta}_2)p(\boldsymbol{\eta}|\mathbf{Z}_p; \boldsymbol{\theta}_2). \qquad (2)$$

Individual examples of such models are introduced in the simulations of Section 5 and applied examples of Section 6.

For the measurement model, we focus on cases where the different items in $\mathbf{Y}$ are conditionally independent of each other given $\boldsymbol{\eta}$ and the measurement model does not depend on $\mathbf{Z}$. These are conventional assumptions which cover most of the latent variable models in common use. For notational convenience we also assume that each item measures only one latent variable in $\boldsymbol{\eta}$. (Generalisations of all of these specifications are discussed briefly in Section 3.) Suppose that $\boldsymbol{\eta} = (\eta_1, \ldots, \eta_J)'$ includes $J \geq 1$ latent variables, and that $\mathbf{Y} = (\mathbf{Y}_1', \ldots, \mathbf{Y}_J')'$ where $\mathbf{Y}_j = (Y_{j1}, \ldots, Y_{jp_j})'$ are the $p_j$ items that measure latent variable $\eta_j$, for $j = 1, \ldots, J$. Suppose further that the measurement models for each $\eta_j$ have distinct parameters, so that $\boldsymbol{\theta}_1 = (\boldsymbol{\theta}_{11}', \ldots, \boldsymbol{\theta}_{1J}')'$ is partititioned similarly. The measurement model is then of the form

$$p(\mathbf{Y}|\boldsymbol{\eta}, \mathbf{Z}; \boldsymbol{\theta}_1) = \prod_{j=1}^{J} p(\mathbf{Y}_j|\eta_j; \boldsymbol{\theta}_{1j}) = \prod_{j=1}^{J} \left[ \prod_{k=1}^{p_j} p(Y_{jk}|\eta_j; \boldsymbol{\theta}_{1j}) \right]. \qquad (3)$$

In other words, the full measurement model of this kind is specified by the collection of the univariate models for the individual items $Y_{jk}$.

Finally, different specifications for the types of the variables define different broad families of latent variable models. For example, if $\mathbf{Y}$, $\boldsymbol{\eta}$ and $\mathbf{Z}_o$ are all taken to be continuous variables and models for them are linear regression models, these models belong to the family of linear structural equation models, for which two-step estimation has been described and implemented by Rosseel & Loh (2024). Here we focus on two other types of models. The first of them is latent trait models (item response theory models) where the measurement items $Y_{jk}$ are categorical variables and the latent variables $\boldsymbol{\eta}$ are continuous (and we further assume that $\boldsymbol{\eta}$ are jointly normally distributed given $\mathbf{Z}_p$). For simplicity, in our examples all the items are binary, with values 0 and 1, and the models for the individual items in (3) are models for Bernoulli-distributed binary outcomes. We then use binary logit models of the form

$$P(Y_{jk} = 1|\eta_j; \boldsymbol{\theta}_{1j}) = \frac{\exp(\tau_{jk} + \lambda_{jk}\eta_j)}{1 + \exp(\tau_{jk} + \lambda_{jk}\eta_j)} \quad (4)$$

and thus $\boldsymbol{\theta}_{1j} = (\tau_{j1}, \lambda_{j1})'$ for $j = 1, \ldots, J; k = 1, \ldots, p_j$.

The second family of models that we consider specifically is latent class models, where both the items and the latent variables are categorical variables. In all of our examples there is only one latent variable in $\boldsymbol{\eta}$ (i.e. $J = 1$), and we denote it by $\eta$ and the measurement items by $\mathbf{Y} = (Y_1, \ldots, Y_p)'$, omitting a redundant subscript. Suppose that $\eta$ has $C$ categories (latent classes) $\eta = 1, \ldots, C$, and the items have categories $Y_k = 1, \ldots, h_k$ for $k = 1, \ldots, p$. The measurement models in (3) are then specified by $p(Y_k|\eta; \boldsymbol{\theta}_1) = \prod_{l=1}^{h_k} P(Y_k = l|\eta = c; \boldsymbol{\theta}_1)^{I(Y_k=l)}$ for $c = 1, \ldots, C$, where $I(\cdot)$ denotes an indicator function. Estimated latent class measurement models are most often summarised by estimates of the item response probabilities $P(Y_k = l|\eta = c; \boldsymbol{\theta}_1)$. For purposes of the variance estimation later, however, we parametrize them in terms of multinomial logistic models

$$P(Y_k = l|\eta = c; \boldsymbol{\theta}_1) = \frac{\exp(\tau_{kl} + \sum_{c=2}^{C} \lambda_{klc}I(\eta = c))}{\sum_{m=1}^{h_k} \exp(\tau_{km} + \sum_{c=2}^{C} \lambda_{kmc}I(\eta = c))}, \quad (5)$$

and thus the measurement parameters $\boldsymbol{\theta}_1$ are $\tau_{kl}$ and $\lambda_{klc}$ for $k = 1, \ldots, p; l = 2, \ldots, h_k$; $c = 2, \ldots, C$ (and those with $l = 1$ are constrained to be 0 for identification).

## 3 Two-step estimation

We focus on likelihood-based estimation of the parameters. Suppose that we have data on $\mathbf{Z}_i$ and $\mathbf{Y}_i$ for units of analysis $i = 1, \ldots, n$, and suppose for now that observations for different units are independent. The log-likelihood for model (1) is then $\ell(\boldsymbol{\theta}) = \ell(\boldsymbol{\theta}_1, \boldsymbol{\theta}_2) =$ $\sum_{i=1}^{n} \log p(\mathbf{Y}_i, \mathbf{Z}_i; \boldsymbol{\theta})$. Maximizing $\ell(\boldsymbol{\theta})$ gives the overall maximum likelihood (ML) estimate of $\boldsymbol{\theta}$. We refer to it as the *one-step estimate* of $\boldsymbol{\theta}$, because it estimates both the measurement parameters $\boldsymbol{\theta}_1$ and the structural parameters $\boldsymbol{\theta}_2$ at once.

*Two-step estimation* proceeds differently. In its step 1, the measurement parameters are estimated from a model where the structural model is simplified as much as is possible while still allowing consistent estimation of $\boldsymbol{\theta}_1$. Let $\ell^*(\boldsymbol{\theta}_1, \boldsymbol{\psi})$ denote the log likelihood for this model, where $\boldsymbol{\psi}$ are parameters of its structural model. This is maximized with respect to the parameters. The estimates $\tilde{\boldsymbol{\theta}}_1$ of $\boldsymbol{\theta}_1$ are taken forward from this step, and the estimates of $\boldsymbol{\psi}$ are discarded. In step 2, we then maximize $\ell(\tilde{\boldsymbol{\theta}}_1, \boldsymbol{\theta}_2)$ with respect to $\boldsymbol{\theta}_2$. This is the same log-likelihood of the full model that would be used for one-step estimation, except that now the measurement parameters $\boldsymbol{\theta}_1$ in it are fixed at their estimated values $\tilde{\boldsymbol{\theta}}_1$ from step 1. Only the structural parameters $\boldsymbol{\theta}_2$ are estimable parameters here. We denote the two-step estimate of them obtained from this step by $\tilde{\boldsymbol{\theta}}_2$.

The specification of $\ell^*(\boldsymbol{\theta}_1, \boldsymbol{\psi})$ in step 1 depends on the form of the overall model (1), in particular the measurement model. Consider first a measurement model of the form (3). From it, the simplest models for step 1 are obtained for each $\eta_j$ separately, by integrating out $\mathbf{Z}$ as well as the other $\eta$s and the items that measure them. This gives

$$\ell^*(\boldsymbol{\theta}_{1j}, \boldsymbol{\psi}) = \sum_{i=1}^{n} \log p(\mathbf{Y}_{ji}; \boldsymbol{\theta}_{1j}, \boldsymbol{\psi}_j) = \sum_{i=1}^{n} \log \int p(\mathbf{Y}_{ji}|\eta_{ji}; \boldsymbol{\theta}_{1j}) p(\eta_{ji}; \boldsymbol{\psi}_j) \, d\eta_{ji} \qquad (6)$$

for $j = 1, \ldots, J$. Here $\boldsymbol{\psi}_j$ are parameters of a univariate distribution for $\eta_{ji}$. For a latent class model this has a multinomial distribution with probabilities $\boldsymbol{\psi} = (\pi_1, \ldots, \pi_C)'$. For a latent trait model we take it to be $p(\eta_{ji}; \boldsymbol{\psi}_k) \sim N(\mu_j^*, \sigma_j^{*2})$, so that $\boldsymbol{\psi}_j = (\mu_j^*, \sigma_j^{*2})'$ (see Kuha & Bakk, 2023 for some further discussion of this choice). In other words, in this situation step 1 entails estimating each $\boldsymbol{\theta}_{1j}$ separately from a model with one latent variable measured by $\mathbf{Y}_j$ and with no covariates, and collecting the resulting estimates as $\tilde{\boldsymbol{\theta}}_1 = (\tilde{\boldsymbol{\theta}}_{11}, \ldots, \tilde{\boldsymbol{\theta}}_{1J}')'$.

If the full measurement model is more complex, the simplest model that allows valid estimation of $\boldsymbol{\theta}_1$ in step 1 needs to be expanded accordingly. For example, if the measurement model includes groups of more than one latent variable that are measured by the same sets of items, each such block of variables should be taken together in step 1. Another possible extension is non-equivalence of measurement, where the measurement model for a $\eta_j$ depends also on some variables $\mathbf{Z}_*$ in $\mathbf{Z}$. In this case the model in step 1 should also be conditional on $\mathbf{Z}^*$ (but still integrated over the rest of $\mathbf{Z}$; see Vermunt & Magidson, 2021, and Lyrvall, Kuha, & Oser, 2025 for more on this topic).

Some estimated variance matrices from these two steps of estimation will also be needed. From step 1 we keep the standard (asymptotic) estimated variance matrix of $\tilde{\boldsymbol{\theta}}_1$, denoted here $\widehat{\text{var}}(\tilde{\boldsymbol{\theta}}_1) = \widehat{\boldsymbol{\Sigma}}_{11}$. If the estimation was done separately for different $\boldsymbol{\theta}_{1j}$, we take $\widehat{\boldsymbol{\Sigma}}_{11}$ to be the block diagonal matrix of the separate estimated variance matrices of $\tilde{\boldsymbol{\theta}}_{1j}$ (see Kuha & Bakk, 2023 for a discussion of this choice). From step 2, what we get as part of standard output is the estimated variance matrix of $\tilde{\boldsymbol{\theta}}_2$ that treats $\tilde{\boldsymbol{\theta}}_1$ as fixed constants rather than estimated paramaters. We denote it by $\widehat{\mathbf{V}}_2$. How these matrices are used as part of the full variance estimation for $\tilde{\boldsymbol{\theta}}_2$ is described in Section 4 below.

If observations of any $\mathbf{Y}_i$ are incomplete, contributions from the missing items are omitted from the likelihoods, but all the observed items are included. Any units $i$ with incomplete data for $\mathbf{Z}_i$ are omitted in step 2. They can still be included in step 1, however, because $\mathbf{Z}_i$ are not used there in any case (unless the measurement model is non-equivalent with respect to them). This is an instance of the more general point that step 1 of two-step estimation can be based on a partially (or even entirely) different sample of data than step 2, as long as we can assume that the same measurement parameters $\boldsymbol{\theta}_1$ hold for both. We note also that if we consider multiple models which have different specifications of the structural model (e.g. different choices of covariates $\mathbf{Z}_p$) but the same measurement model, only step 2 needs to be re-done for each of them while step 1 only needs to be carried out once.

We have assumed here that the data consist of $n$ independent observations. Other dependence structures are also possible, such as hierarchical (multilevel) data where lower-level units are clustered within higher-level ones (e.g. school children within classes). In this case the log-likelihood is a sum over the highest-level clusters and the structural model may also include hierarchical models for the latent variables. One such situation is considered by Di Mari et al. (2023), who describe two-step methods for multilevel latent class models. The principle of two-step estimation is unchanged by this, as is the method of variance estimation that we describe in Section 4.2.

# 4 Variance estimation for two-step estimates

## 4.1 Asymptotic variance estimation

Asymptotic properties of the two-step estimator $\tilde{\boldsymbol{\theta}}_2$ of the parameters of the structural model follow the general theory of pseudo maximum likelihood estimators of Gong & Samaniego (1981). They are described in more detail in Bakk & Kuha (2018) and Kuha & Bakk (2023). In short, $\tilde{\boldsymbol{\theta}}_2$ is consistent for $\boldsymbol{\theta}_2$ and asymptotically normally distributed under general regularity conditions. What concerns us here is the form and estimation of its asymptotic variance(-covariance) matrix.

Denote the Fisher information matrix for $\boldsymbol{\theta}$ in the full model (1) by

$$\boldsymbol{\mathcal{I}}(\boldsymbol{\theta}^*) = \sum_{i=1}^{n} \boldsymbol{\mathcal{I}}_i(\boldsymbol{\theta}^*) = \begin{bmatrix} \boldsymbol{\mathcal{I}}_{11} & \\ \boldsymbol{\mathcal{I}}'_{12} & \boldsymbol{\mathcal{I}}_{22} \end{bmatrix}$$

where $\boldsymbol{\theta}^*$ denotes the true value of $\boldsymbol{\theta}$ and the partitioning corresponds to $\boldsymbol{\theta}_1$ and $\boldsymbol{\theta}_2$. The asymptotic variance matrix of the one-step ML estimate of $\boldsymbol{\theta}$ is thus $\boldsymbol{\mathcal{I}}^{-1}(\boldsymbol{\theta}^*)$. Let $\boldsymbol{\Sigma}_{11}$ be the asymptotic variance matrix of the step-1 estimator $\tilde{\boldsymbol{\theta}}_1$. The asymptotic variance matrix of the two-step estimator $\tilde{\boldsymbol{\theta}}_2$ is then

$$\mathbf{V} = \boldsymbol{\mathcal{I}}_{22}^{-1} + \boldsymbol{\mathcal{I}}_{22}^{-1}\boldsymbol{\mathcal{I}}_{12}'\boldsymbol{\Sigma}_{11}\boldsymbol{\mathcal{I}}_{12}\boldsymbol{\mathcal{I}}_{22}^{-1} \equiv \mathbf{V}_2 + \mathbf{V}_1. \tag{7}$$

In essence, $\mathbf{V}_2$ describes the variability in $\tilde{\boldsymbol{\theta}}_2$ given a fixed value of the step-1 estimate $\tilde{\boldsymbol{\theta}}_1$, and $\mathbf{V}_1$, accounts for the additional variability arising from the fact that $\boldsymbol{\theta}_1$ are not known but rather estimated by $\tilde{\boldsymbol{\theta}}_1$ with their own sampling variability. Previous simulation studies show, as is also expected on theoretical grounds, that estimated standard errors based on (7) perform very well when the sample sizes are not too small.

Estimates of $\boldsymbol{\Sigma}_{11}$ and $\mathbf{V}_2 = \boldsymbol{\mathcal{I}}_{22}^{-1}$ are obtained from steps 1 and 2 of estimation respectively, as explained in Section 3. This does not, however, give us an estimate of the cross-derivative matrix $\boldsymbol{\mathcal{I}}_{12}$. It is part of the information matrix for the joint model (1) but should be evaluated at the two-step estimates $\tilde{\boldsymbol{\theta}} = (\tilde{\boldsymbol{\theta}}_1', \tilde{\boldsymbol{\theta}}_2')'$. It is thus the only element of $\mathbf{V}$ which does not come naturally from either of the steps of two-step estimation but which will need to be calculated separately. This may often make estimation of $\boldsymbol{\mathcal{I}}_{12}$ the most demanding and inconvenient element of the whole two-step procedure. For linearr structural equation models it is conveniently implemented in the *lavaan* package in R (Rosseel, 2012). For other types of models — including the ones considered in this paper — it will require additional derivations and/or programming. One option is to derive and code the necessary derivatives in a closed form, but this is model-specific and unlikely to be feasible in general (see Skrondal & Kuha, 2012, for an example of it). A more generally applicable approach involves a further call to estimation software that was also used for step 2, relying on its procedures for evaluating information matrices. Here the code is as for one-step estimation of the model, $\tilde{\boldsymbol{\theta}}$ are used as the starting values, the estimation algorithm is interrupted before or after one iteration, and an estimate of $\boldsymbol{\mathcal{I}}_{12}$ is extracted from the estimate of $\boldsymbol{\mathcal{I}}$ at that point. This method was employed by Bakk & Kuha (2018) for latent class models (using the Latent GOLD software) and by Kuha & Bakk (2023) for latent trait models (using the Mplus software).

## 4.2 Partially simulation-based variance estimation

Let $\widehat{\mathbf{V}} = \widehat{\mathbf{V}}_2 + \widehat{\mathbf{V}}_1$ denote the estimate of the variance matrix $\mathbf{V} = \text{var}(\tilde{\boldsymbol{\theta}}_2)$ that is obtained as described in Section 4.1. What we propose here is an alternative way of estimating $\mathbf{V}$, using an approach which employs simulation to estimate the matrix $\mathbf{V}_1$. This avoids the need for any separate closed-form or numerical evaluation of the cross-derivative matrix $\boldsymbol{\mathcal{I}}_{12}$.

Our approach is motivated by an ostensibly different two-part expression of the variance matrix, namely the standard decomposition of it according to the law of total variance as

$$\text{var}(\tilde{\boldsymbol{\theta}}_2) = \text{E}_{\tilde{\boldsymbol{\theta}}_1}[\text{var}(\tilde{\boldsymbol{\theta}}_2|\tilde{\boldsymbol{\theta}}_1)] + \text{var}_{\tilde{\boldsymbol{\theta}}_1}[\text{E}(\tilde{\boldsymbol{\theta}}_2|\tilde{\boldsymbol{\theta}}_1)] \tag{8}$$

where the outer expectation and variance are over the sampling distribution of the step-1 estimator $\tilde{\boldsymbol{\theta}}_1$, treated as a random variable.

We estimate the first term in (8) by an estimate of $\text{var}(\tilde{\boldsymbol{\theta}}_2|\tilde{\boldsymbol{\theta}}_1)$ evaluated at the point estimate $\tilde{\boldsymbol{\theta}}_1$. In other words, this is the same $\widehat{\mathbf{V}}_2 = \widehat{\boldsymbol{\mathcal{I}}}_{22}^{-1}$ that we would use to estimate $\mathbf{V}_2$ in the the standard asymptotic expression (7).

The difference between the two approaches is in the second term of the variance expressions. We propose to use straightforward Monte Carlo simulation over the estimated sampling distribution of $\tilde{\boldsymbol{\theta}}_1$ to estimate the second term of (8). It proceeds as follows:

(i) Generate $\tilde{\boldsymbol{\theta}}_1^{(j)}$ for $j = 1, \ldots, M$, independently from $N(\tilde{\boldsymbol{\theta}}_1, \widehat{\boldsymbol{\Sigma}}_{11})$, the estimated asymptotic distribution of $\tilde{\boldsymbol{\theta}}_1$ based on step 1 of two-step estimation.

(ii) Repeat step 2 of two-step estimation with $\boldsymbol{\theta}_1$ fixed at each $\tilde{\boldsymbol{\theta}}_1^{(j)}$ in turn. This gives estimates $\tilde{\boldsymbol{\theta}}_2^{(j)}$ of $\boldsymbol{\theta}_2$, for $j = 1, \ldots, M$.

(iii) Estimate $\text{var}_{\tilde{\boldsymbol{\theta}}_1}[\text{E}(\tilde{\boldsymbol{\theta}}_2|\tilde{\boldsymbol{\theta}}_1)]$ in (8) by the sample variance matrix of $\tilde{\boldsymbol{\theta}}_2^{(1)},\ldots,\tilde{\boldsymbol{\theta}}_2^{(M)}$. We denote this estimate by $\widehat{\mathbf{V}}_{1M}$.

The simulation-based estimate of the variance matrix of $\tilde{\boldsymbol{\theta}}_2$ is then $\widehat{\mathbf{V}}_M = \widehat{\mathbf{V}}_2 + \widehat{\mathbf{V}}_{1M}$.

At this point the relationship between the asymptotic and the simulation-based variance estimates is not obvious, because the two decompositions (7) and (8) are different. We have seen that it is natural to estimate the first term of both by $\widehat{\mathbf{V}}_2$, but any connection between $\widehat{\mathbf{V}}_1$ and $\widehat{\mathbf{V}}_{1M}$ is not immediately apparent. However, there is such a connection, which shows that $\widehat{\mathbf{V}}_{1M}$ is also a simulation-based estimate of $\mathbf{V}_1$ in the asymptotic variance expression (7). To see this, consider first the asymptotic distribution of the joint (one-step) ML estimator $\hat{\boldsymbol{\theta}} = (\hat{\boldsymbol{\theta}}_1', \hat{\boldsymbol{\theta}}_2')'$ of $\boldsymbol{\theta} = (\boldsymbol{\theta}_1', \boldsymbol{\theta}_2')'$. This is multivariate normal, as $\hat{\boldsymbol{\theta}} \sim N(\boldsymbol{\theta}^*, \boldsymbol{\Sigma}^*)$ with the variance matrix $\boldsymbol{\Sigma}^* = \mathcal{I}^{-1}(\boldsymbol{\theta}^*) = \mathcal{I}^{-1}$, which we partition as

$$\boldsymbol{\Sigma}^* = \begin{bmatrix} \boldsymbol{\Sigma}_{11}^* & \\ \boldsymbol{\Sigma}_{12}^{*\prime} & \boldsymbol{\Sigma}_{22}^* \end{bmatrix}.$$

Consider now the conditional distribution obtained from this for the estimate of $\boldsymbol{\theta}_2$ given a fixed value of the estimate of $\boldsymbol{\theta}_1$. Although this is derived as the conditional distribution of $\hat{\boldsymbol{\theta}}_2$ given $\hat{\boldsymbol{\theta}}_1$, it also applies to two-step estimates as they were defined in Section 3 above. This is because step 2 is then equal to ML estimation of $\boldsymbol{\theta}_2$ when $\boldsymbol{\theta}_1 = \tilde{\boldsymbol{\theta}}_1$ is treated as a known quantity. We thus use the same asymptotic conditional distribution derived from this also for $\tilde{\boldsymbol{\theta}}_2$ given a fixed value of $\tilde{\boldsymbol{\theta}}_1$. It is multivariate normal, with mean and variance matrix

$$\text{E}(\tilde{\boldsymbol{\theta}}_2|\tilde{\boldsymbol{\theta}}_1) = \text{E}(\tilde{\boldsymbol{\theta}}_2) - \boldsymbol{\Sigma}_{12}^{*\prime}\boldsymbol{\Sigma}_{11}^{*-1}(\tilde{\boldsymbol{\theta}}_1 - \text{E}(\tilde{\boldsymbol{\theta}}_1)) = \boldsymbol{\theta}_2^* - \boldsymbol{\Sigma}_{12}^{*\prime}\boldsymbol{\Sigma}_{11}^{*-1}(\tilde{\boldsymbol{\theta}}_1 - \boldsymbol{\theta}_1^*) \quad \text{and} \quad (9)$$

$$\text{var}(\tilde{\boldsymbol{\theta}}_2|\tilde{\boldsymbol{\theta}}_1) = \boldsymbol{\Sigma}_{22}^* - \boldsymbol{\Sigma}_{12}^{*\prime}\boldsymbol{\Sigma}_{11}^{*-1}\boldsymbol{\Sigma}_{12}^* \quad (10)$$

where the last expression in (9) follows because $\tilde{\boldsymbol{\theta}}_1$ and $\tilde{\boldsymbol{\theta}}_2$ are consistent for $\boldsymbol{\theta}_1$ and $\boldsymbol{\theta}_2$ respectively.

Consider first (10) over the asymptotic distribution of $\tilde{\boldsymbol{\theta}}_1$, to obtain an asymptotic expression for the first term of (8) and to see how this matches the first term of (7). Note first that (10) is not a function of $\tilde{\boldsymbol{\theta}}_1$, so $\mathrm{E}_{\tilde{\boldsymbol{\theta}}_1}[\mathrm{var}(\tilde{\boldsymbol{\theta}}_2|\tilde{\boldsymbol{\theta}}_1)] = \mathrm{var}(\tilde{\boldsymbol{\theta}}_2|\tilde{\boldsymbol{\theta}}_1)$. Next, the key to the connection is the fact that $\boldsymbol{\Sigma}^* = \boldsymbol{\mathcal{I}}^{-1}$. Standard results for inverses of partitioned matrices then show that the (2,2) block of $\boldsymbol{\Sigma}^{*-1}$, i.e. $\boldsymbol{\mathcal{I}}_{22}$, is $(\boldsymbol{\Sigma}^*_{22} - \boldsymbol{\Sigma}^{*\prime}_{12}\boldsymbol{\Sigma}^{*-1}_{11}\boldsymbol{\Sigma}^*_{12})^{-1}$, and thus $\mathrm{var}(\tilde{\boldsymbol{\theta}}_2|\tilde{\boldsymbol{\theta}}_1) = \boldsymbol{\mathcal{I}}^{-1}_{22}$. The first terms of (7) and (8) are thus asymptotically the same, and both are estimated by $\widehat{\mathbf{V}}_2 = \widehat{\boldsymbol{\mathcal{I}}}^{-1}_{22}$ from step 2 of two-step estimation.

For the second term of (8), we get from (9)

$$\mathrm{var}_{\tilde{\boldsymbol{\theta}}_1}[\mathrm{E}(\tilde{\boldsymbol{\theta}}_2|\tilde{\boldsymbol{\theta}}_1)] = \boldsymbol{\Sigma}^{*\prime}_{12}\boldsymbol{\Sigma}^{*-1}_{11}\ \mathrm{var}(\tilde{\boldsymbol{\theta}}_1)\ \boldsymbol{\Sigma}^{*-1}_{11}\boldsymbol{\Sigma}^*_{12}.$$

On the other hand, using results for partitioned matrices again we have the expression for $\boldsymbol{\mathcal{I}}_{22}$ above and $\boldsymbol{\mathcal{I}}'_{12} = -(\boldsymbol{\Sigma}^*_{22} - \boldsymbol{\Sigma}^{*\prime}_{12}\boldsymbol{\Sigma}^{*-1}_{11}\boldsymbol{\Sigma}^*_{12})^{-1}\,\boldsymbol{\Sigma}^{*\prime}_{12}(\boldsymbol{\Sigma}^*_{11})^{-1}$, so that $\boldsymbol{\mathcal{I}}^{-1}_{22}\boldsymbol{\mathcal{I}}'_{12} = -\boldsymbol{\Sigma}^{*\prime}_{12}\boldsymbol{\Sigma}^{*-1}_{11}$ and

$$\mathrm{var}_{\tilde{\boldsymbol{\theta}}_1}[\mathrm{E}(\tilde{\boldsymbol{\theta}}_2|\tilde{\boldsymbol{\theta}}_1)] = \boldsymbol{\mathcal{I}}^{-1}_{22}\,\boldsymbol{\mathcal{I}}'_{12}\ \mathrm{var}(\tilde{\boldsymbol{\theta}}_1)\,\boldsymbol{\mathcal{I}}_{12}\,\boldsymbol{\mathcal{I}}^{-1}_{22} \qquad (11)$$

where $\mathrm{var}(\tilde{\boldsymbol{\theta}}_1)$ is $\boldsymbol{\Sigma}_{11}$ in the notation of Section 4.1. Therefore the second terms of (7) and (8) are asymptotically equal. The simulation-based approach described in this section allows us to estimate this term without having to evaluate $\boldsymbol{\mathcal{I}}_{12}$ explicitly.

Because this procedure eliminates any separate steps to estimate $\boldsymbol{\mathcal{I}}_{12}$, it restores the full separation of steps 1 and 2 which is an appealing feature of two-step estimation. From step 1 we now save not only the estimates $\tilde{\boldsymbol{\theta}}_1$ and $\widehat{\boldsymbol{\Sigma}}_{11}$ but also the simulation draws $\tilde{\boldsymbol{\theta}}^{(1)}_1, \ldots, \tilde{\boldsymbol{\theta}}^{(M)}_1$. They can then be used in step 2 for point and variance estimation of any structural model which is combined with the measurement model with the same parameters $\boldsymbol{\theta}_1$.

Computer implementation of this combination of two-step estimation and simulation-based variance estimation requires software that allows for one-step estimation with fixed values for the measurement parameters, together with practicable ways of managing the process of repeatedly changing those fixed values and assembling the results. In the supplementary materials of this article we provide code for doing this with a combination of the R and Mplus software, as described in Section 6. The additional computational cost of the simulations is repeating step-2 estimation for a further $M$ times. The computing time may thus be expected to be roughly $M$ times that of two-step estimation without the simulation-based variance estimation.

# 5 Simulation studies

Here we use simulation studies to examine how our proposed, partially simulation-based method performs in estimating standard errors of two-step estimates of structural parameters. As a benchmark, we compare it to standard errors of two-step and one-step estimators from closed-form asymptotic variance formulas, the properties of which are well established. We also examine how the performance of our method depends on the number of simulation draws ($M$) used for it. Section 5.1 reports simulations for latent trait models, and Section 5.2 for latent class models.

## 5.1 Latent trait models

In the first simulation study we examine latent trait models with continuous latent variables and binary measurement items. The simulation settings are based on those of the first simulation study of Kuha & Bakk (2023, Section 4.1, Tables 1 and 2). In the notation of Section 2, the structural model is such that there is no $\mathbf{Z}$ and there are two latent traits $\eta = (\eta_1, \eta_2)'$, with $\eta_1$ treated as an explanatory variable for $\eta_2$. Specifically, for units $i = 1, \ldots, n$, $\eta_i$ are generated as $\eta_{1i} \sim N(0, 1)$ and $\eta_{2i}|\eta_{1i} \sim N(\beta_0 + \beta_1 \eta_{1i}, \sigma_2^2)$ with $\beta_0 = 0$. Thus marginally $\eta_{2i} \sim N(0, \beta_1^2 + \sigma^2)$, and the $R^2$ statistic for $\eta_{i2}$ given $\eta_{1i}$ is $R_\eta^2 = \beta_1^2/(\beta_1^2 + \sigma_2^2)$. The parameters $(\beta_1, \sigma_2^2)$ are set so that $\beta_1^2 + \sigma^2 = \text{var}(\eta_{2i}) = 1$ and $R_\eta^2$ takes on specified values as described below.

The latent variables $\eta_{1i}$ and $\eta_{2i}$ are each measured by $p=4$ distinct binary measurement items $Y_{ikj}$ for $i=1,\ldots,n$, $k=1,2$, $j=1,\ldots,p$, with values 0 and 1. All items are conditionally independent of each other given the latent variables, and the measurement models are of the logistic form $\text{logit}[P(Y_{ikj}=1|\eta_{ik})]=\tau+\lambda\eta_{ik}$. In the data-generating process for a given simulation setting all $2p$ items have the same true value of $\tau$ and the same $\lambda$, but in the model estimation they are each estimated as distinct parameters, as in equation (4) above. The loading parameter $\lambda$ is set with reference to the linear model for a notional continuous latent variable $Y_{ikj}^*$ which implies the logistic model for $Y_{ikj}$. Here $R^2$ for $Y_{ikj}^*$ given $\eta_{ik}$ is $R_Y^2 = \lambda^2 \text{var}(\eta_{ik})/(\lambda^2 \text{var}(\eta_{ik}) + \pi^2/3) = \lambda^2/(\lambda^2+\pi^2/3)$, and $\lambda$ is set to obtain specified values of $R_Y^2$. The measurement intercept $\tau$ is set so that the marginal probability $P(Y_{ikj}=1)$ is 0.5 for all items.

We focus on results for the estimates of $\beta_1$, and consider all 18 combinations of $n = 500, 1000, 2000$, $R_\eta^2 = 0, 0.2, 0.4$ (corresponding to $\beta_1 = 0, 0.447, 0.632$ respectively), and $R_Y^2 = 0.4, 0.6$. The results in Kuha & Bakk (2023) indicate that these are settings, with reasonably large sample sizes and sufficiently strong measurement models, where point estimates of $\beta_1$ and their estimated standard errors based on the asymptotic formulas behave generally well. We consider such situation here in order to focus on the comparative performance of our proposed method of standard error estimation, without distractions from other possible errors of estimation. For each setting, 250 simulated datasets were generated. The computations were carried out using a combination of Mplus 6.12 (Muthén & Muthén, 2010) and R software (R Core Team, 2024), as in Kuha & Bakk (2023), here extended to include also our simulation-based method of variance estimation (this computing set-up is descibed in more detail in the application Section 6 below).

We examine the performance of estimated standard errors of estimates of $\beta_1$ and con-fidence intervals which use them. The following estimates are considered: for two-step estimates naive variance estimates which use only $\mathbf{V}_2$ in (7), correct asymptotic variance estimates based on all of (7), and our proposed simulation-based approach, with $M = 50, 100, 200, 500, 1000$. For comparison, we also report results for one-step estimates of $\beta_1$, using standard asymptotic variance estimates. In some simulations one-step estimation gave extreme estimates of $\beta_1$, or the standard errors for them failed to be calculated. This may be because the estimation algorithm converged to local maxima of the log likelihood. To avoid this distorting the comparisons, we omit such simulations from the results for all methods (using the ad hoc criterion that a simulation is omitted if the one-step standard error failed or if the one- and two-step point estimates of $\beta_1$ differed by more than 0.3 in absolute value). Because of this, the results reported below omit between 0 and 14 of the 250 simulated datasets for different simulation settings. The results are summarised in two ways. Table 1 shows, for each standard error estimator, the ratio between the average estimated standard error across the simulations and the simulation standard deviation of the corresponding point estimates of $\beta_1$. These ratios should be as close to 1 as possible. Table 2 then shows the simulation coverage of 95% confidence intervals obtained using the point estimates and the different standard error estimates.

The simulation results show that our proposed method works well. We note first that the naive standard errors of two-step estimates underestimate the true variability of the point estimates, often severely so, when the structural coefficient $\beta_1$ is not zero. Realised coverage of the confidence intervals is then clearly below the nominal 95%. On the other hand, all the approaches which account for the uncertainty from step 1 of two-step estimates correct this underestimation and achieve approximately correct coverage. They also perform very similarly. The proposed simulation-based approach gives essentially the same results as the full asymptotic variance formula. Furthermore, its performance is essentially unaffected by the number of simulation draws $M$. In Table 1 the ratios (i.e. in effect the simulation averages of the standard errors) are almost the same from $M = 50$ to $M = 1000$. The simulation variability of these estimates (which is not shown here) is larger when $M$ smaller, which could in principle affect the coverage of the confidence intervals. Table 2 shows, however, that this does not happen, and small values of $M$ again achieve as good a coverage as larger ones. Finally, all these (asymptotic and simulation-based) standard errors for the two-step approach which account for the first-step variation perform as well as the asymptotic standard error estimates for the one-step estimates, and in some cases somewhat better.

## 5.2 Latent class models

The second simulation study considers latent class models with a single latent-class variable $\eta$ with $C = 3$ categories. The structural model includes one observed continuous explanatory variable $Z$ for $\eta$, generated as $Z_i \sim N(0, 1)$ for units $i = 1, \ldots, n$. The structural model for $\eta_i$ given $Z_i$ is parametrized as the multinomial logistic model

$$P(\eta_i = c|Z_i) = \frac{\exp(\beta_{0c} + \beta_{1c}Z_i)}{\sum_{g=1}^{C} \exp(\beta_{0g} + \beta_{1g}Z_i)}, \qquad (12)$$

where $\beta_{01} = \beta_{11} = 0$ for identification, and $\beta_{02} = 0.1$ and $\beta_{03} = 0.2$. We set $\beta_{12}$ and $\beta_{13}$ to be equal, and consider three simulation settings, with $\beta_{12} = \beta_{13} = 0$, $0.447$, or $0.632$. We refer to these settings as no effect, mild effect, and strong effect of $Z_i$ on $\eta_i$ in the structural model. Averaged over the distribution of $Z_i$, this implies marginal proportions of the three latent classes between $(0.32, 0.33, 0.35)$ in the no-effect case and $(0.28, 0.33, 0.39)$ in the strong-effect case.

The latent variable $\eta_i$ is measured by $p = 10$ binary measurement items $Y_{ij}$ for $i = 1, \ldots, n$, $j = 1, \ldots, p$, each with values 0 and 1. Items are conditionally independent of each other given the latent class variable. We vary the strength of the measurement model, in terms of class separation, by allowing the item-response probabilities for the most likely response to be either 0.87 or 0.91, corresponding to 0.8 (moderate class separation) and 0.9 (large class separation) of Magidson (1981)'s entropy-based $R^2$. We refer to this measure also as $R^2_Y$ in the results Tables 3 and 4 below.

In the data-generating process for a given simulation setting, all items use either the moderate or the large class separation setting, but in the estimation all measurement parameters are estimated separately. The measurement models for different latent classes differ in which values have the high probability for each of the ten items. For the first class it is value 1 for all items, for the second class 1 for five items and 0 for the other five, and for the third class 0 for all items.

We consider all 18 combinations of No, Mild and Strong covariate effect in the structural model, Moderate and Large class separation in the measurement model, and sample sizes $n = 500$, 1000 and 2000, and generate 250 simulated datasets for each setting. Data generation and model estimation were carried out in R (Venables et al., 2013), extending the functionalities available in the package `multilevLCA` (Lyrvall, Di Mari, et al., 2025). Here again calculation of one-step estimates or their standard errors failed in a small number of cases (65 out of the $18 \times 250$ total samples), and these cases are omitted from the results for all estimates (using the ad hoc criterion that a simulation is omitted if the one-step standard error failed or if the one- and two-step point estimates of either structrural coefficient differed by more than 0.5 in absolute value). We consider the same estimators and the same quantities of interest as in the simulations of Section 5.1. Here there are two structural parameters of interest, the coefficients $\beta_{12}$ and $\beta_{13}$ in (12), and the results below refer to the estimates for $\beta_{12}$

Simulation results for the latent class models are shown in Table 3 for the accuracy of the estimated standard errors and in Table 4 for the coverage of 95% confidence intervals. The conclusions are qualitatively the same as in the latent-trait simulations in Section 5.1. Our simulation-based variance estimation performs as well as asymptotic variance estimates and is unaffected by the number ($M$) of simulation draws used for it. Here the latter result is very clear, as the results are the same for all $M$ (except for the naive estimates which use $M = 0$) up to the three decimal places shown in Tables 3 and 4. This is largely due to the fact that here the contribution to the variances from step 1 of the estimation is small. The simulation-based variance estimation correctly estimates the size of that contribution, already with small values of $M$.

Table 1: Simulation results for different standard error estimates of estimated structural regression coefficient $\beta_1$ of a latent covariate $\eta_1$ for a conditionally normally distributed latent response $\eta_2$. For each estimate, the table shows the ratio between the average estimated standard error across the simulations and the simulation standard deviation of the corresponding point estimates of $\beta_1$.

| | | | One-step | Two-step | | | | | | |
|---|---|---|---|---|---|---|---|---|---|---|
| | | | | | | \multicolumn{5}{c}{Simulation-based, with $M$ simulation draws: $M =$} | | | | |
| $R^2_\eta$ | $n$ | $R^2_Y$ | asympt. | naive | asympt. | 50 | 100 | 200 | 500 | 1000 |
| 0 | 500 | 0.4 | 0.951 | 0.934 | 0.963 | 0.968 | 0.969 | 0.969 | 0.968 | 0.968 |
| | | 0.6 | 0.999 | 0.990 | 1.008 | 1.012 | 1.012 | 1.012 | 1.012 | 1.012 |
| | 1000 | 0.4 | 1.011 | 1.005 | 1.021 | 1.023 | 1.023 | 1.023 | 1.022 | 1.022 |
| | | 0.6 | 1.023 | 1.026 | 1.035 | 1.036 | 1.036 | 1.036 | 1.036 | 1.036 |
| | 2000 | 0.4 | 1.062 | 1.058 | 1.066 | 1.066 | 1.066 | 1.066 | 1.066 | 1.066 |
| | | 0.6 | 1.737 | 0.951 | 0.956 | 0.956 | 0.956 | 0.956 | 0.956 | 0.956 |
| 0.2 | 500 | 0.4 | 1.004 | 0.643 | 1.003 | 1.039 | 1.045 | 1.050 | 1.049 | 1.050 |
| | | 0.6 | 0.986 | 0.653 | 0.995 | 1.030 | 1.039 | 1.038 | 1.041 | 1.042 |
| | 1000 | 0.4 | 1.001 | 0.656 | 1.014 | 1.029 | 1.031 | 1.033 | 1.036 | 1.035 |
| | | 0.6 | 0.999 | 0.677 | 1.022 | 1.040 | 1.040 | 1.043 | 1.044 | 1.045 |
| | 2000 | 0.4 | 0.890 | 0.601 | 0.933 | 0.939 | 0.939 | 0.940 | 0.941 | 0.941 |
| | | 0.6 | 0.967 | 0.657 | 0.994 | 1.005 | 1.010 | 1.010 | 1.009 | 1.008 |
| 0.4 | 500 | 0.4 | 1.103 | 0.558 | 1.041 | 1.087 | 1.086 | 1.092 | 1.086 | 1.088 |
| | | 0.6 | 0.956 | 0.508 | 0.937 | 0.991 | 0.991 | 0.986 | 0.988 | 0.989 |
| | 1000 | 0.4 | 0.999 | 0.549 | 1.027 | 1.042 | 1.045 | 1.044 | 1.044 | 1.046 |
| | | 0.6 | 1.046 | 0.578 | 1.048 | 1.060 | 1.064 | 1.070 | 1.073 | 1.074 |
| | 2000 | 0.4 | 1.102 | 0.583 | 1.083 | 1.096 | 1.090 | 1.093 | 1.094 | 1.094 |
| | | 0.6 | 0.988 | 0.566 | 1.021 | 1.032 | 1.031 | 1.030 | 1.033 | 1.032 |

Note: $R^2_Y$ and $R^2_\eta$ denote the $R^2$ statistics in models for $Y_j$ and $\eta_2$ respectively, as explained in the text.

Table 2: Simulation results for different standard error estimates of estimated structural regression coefficient $\beta_1$ of a latent covariate $\eta_1$ for a conditionally normally distributed latent response $\eta_2$. The table shows the simulation coverage of 95% confidence intervals for $\beta_1$ calculated using each standard error estimate and the corresponding point estimate of $\beta_1$.

|  |  |  | One-step | Two-step | | | | | | |
|---|---|---|---|---|---|---|---|---|---|---|
|  |  |  |  |  |  | Simulation-based, with $M$ simulation draws: $M =$ | | | | |
| $R^2_\eta$ | $n$ | $R^2_Y$ | asympt. | naive | asympt. | 50 | 100 | 200 | 500 | 1000 |
| 0 | 500 | 0.4 | 96.8 | 94.4 | 96.8 | 97.2 | 97.2 | 97.2 | 96.8 | 96.8 |
|  |  | 0.6 | 96.0 | 95.6 | 96.0 | 96.4 | 96.4 | 96.8 | 96.8 | 96.8 |
|  | 1000 | 0.4 | 96.0 | 94.4 | 96.0 | 96.0 | 96.4 | 96.0 | 96.0 | 96.0 |
|  |  | 0.6 | 98.0 | 97.2 | 98.0 | 98.0 | 98.0 | 98.0 | 98.0 | 98.0 |
|  | 2000 | 0.4 | 96.8 | 96.4 | 96.8 | 97.2 | 97.2 | 97.2 | 97.2 | 96.8 |
|  |  | 0.6 | 95.9 | 95.9 | 95.9 | 95.9 | 95.9 | 95.9 | 95.9 | 95.9 |
| 0.2 | 500 | 0.4 | 94.8 | 76.8 | 94.4 | 94.8 | 94.4 | 94.8 | 94.8 | 94.8 |
|  |  | 0.6 | 94.2 | 79.0 | 94.2 | 93.8 | 95.1 | 95.5 | 95.5 | 95.5 |
|  | 1000 | 0.4 | 94.4 | 81.6 | 94.4 | 94.8 | 95.2 | 95.2 | 95.2 | 94.8 |
|  |  | 0.6 | 94.6 | 82.2 | 96.3 | 96.7 | 96.3 | 96.7 | 96.7 | 96.7 |
|  | 2000 | 0.4 | 89.5 | 74.9 | 91.1 | 90.3 | 91.5 | 91.1 | 91.1 | 91.5 |
|  |  | 0.6 | 94.6 | 81.2 | 95.8 | 95.4 | 95.8 | 95.8 | 96.2 | 96.2 |
| 0.4 | 500 | 0.4 | 95.6 | 70.4 | 93.6 | 94.0 | 94.4 | 94.0 | 94.0 | 94.0 |
|  |  | 0.6 | 90.9 | 70.4 | 93.0 | 92.6 | 93.4 | 93.8 | 93.0 | 93.4 |
|  | 1000 | 0.4 | 95.6 | 70.8 | 96.0 | 95.6 | 95.2 | 95.6 | 95.6 | 95.6 |
|  |  | 0.6 | 97.0 | 73.7 | 97.9 | 97.5 | 97.9 | 97.9 | 97.9 | 97.9 |
|  | 2000 | 0.4 | 96.8 | 75.6 | 96.8 | 97.2 | 96.4 | 96.8 | 97.2 | 96.8 |
|  |  | 0.6 | 94.6 | 76.9 | 94.6 | 94.6 | 93.8 | 94.2 | 94.6 | 94.6 |

Note: $R^2_Y$ and $R^2_\eta$ denote the $R^2$ statistics in models for $Y_j$ and $\eta_2$ respectively, as explained in the text.

Table 3: Simulation results for different standard error estimates of estimated structural regression coefficients $\beta_{12}$ of a continuous covariate $Z$ in a model for a discrete latent class variable $\eta$. For each setting, the table shows the ratio between the average estimated standard error accross the simulations and the simulation standard deviation of the corresponding point estimates of $\beta_{12}$.

| Structural covariance effect | $n$ | $R_Y^2$ | One-step asympt. | Two-step | | | | | | |
|---|---|---|---|---|---|---|---|---|---|---|
| | | | | naive | asympt. | Simulation-based, with $M$ simulation draws: $M =$ | | | | |
| | | | | | | 50 | 100 | 200 | 500 | 1000 |
| No effect | 500 | 0.8 | 0.977 | 0.990 | 1.002 | 0.999 | 0.999 | 0.999 | 0.999 | 0.999 |
| | | 0.9 | 1.043 | 1.064 | 1.073 | 1.066 | 1.066 | 1.066 | 1.066 | 1.066 |
| | 1000 | 0.8 | 0.960 | 0.974 | 0.978 | 0.977 | 0.977 | 0.977 | 0.977 | 0.977 |
| | | 0.9 | 1.023 | 1.039 | 1.041 | 1.039 | 1.039 | 1.039 | 1.039 | 1.039 |
| | 2000 | 0.8 | 1.007 | 1.010 | 1.011 | 1.011 | 1.011 | 1.011 | 1.011 | 1.011 |
| | | 0.9 | 0.925 | 0.925 | 0.926 | 0.925 | 0.925 | 0.925 | 0.925 | 0.925 |
| Mild effect | 500 | 0.8 | 1.003 | 1.074 | 1.088 | 1.091 | 1.091 | 1.091 | 1.091 | 1.091 |
| | | 0.9 | 1.010 | 1.014 | 1.022 | 1.016 | 1.016 | 1.016 | 1.016 | 1.016 |
| | 1000 | 0.8 | 1.009 | 1.011 | 1.016 | 1.018 | 1.018 | 1.018 | 1.018 | 1.018 |
| | | 0.9 | 1.024 | 1.019 | 1.022 | 1.020 | 1.020 | 1.020 | 1.020 | 1.020 |
| | 2000 | 0.8 | 1.018 | 1.022 | 1.024 | 1.025 | 1.025 | 1.025 | 1.025 | 1.025 |
| | | 0.9 | 1.058 | 1.059 | 1.059 | 1.059 | 1.059 | 1.059 | 1.059 | 1.059 |
| Strong effect | 500 | 0.8 | 0.983 | 0.997 | 1.014 | 1.025 | 1.026 | 1.026 | 1.026 | 1.026 |
| | | 0.9 | 1.050 | 1.059 | 1.069 | 1.064 | 1.064 | 1.064 | 1.064 | 1.064 |
| | 1000 | 0.8 | 1.008 | 1.009 | 1.016 | 1.023 | 1.023 | 1.023 | 1.023 | 1.023 |
| | | 0.9 | 0.957 | 0.962 | 0.965 | 0.964 | 0.964 | 0.964 | 0.964 | 0.964 |
| | 2000 | 0.8 | 1.028 | 1.029 | 1.031 | 1.036 | 1.036 | 1.036 | 1.036 | 1.036 |
| | | 0.9 | 0.977 | 0.979 | 0.980 | 0.980 | 0.980 | 0.980 | 0.980 | 0.980 |

Note: Sizes of the covariate effects and the $R_Y^2$ statistic for items $Y_j$ are defined in the text.

Table 4: Simulation results for different standard error estimates of estimated structural regression coefficients $\beta_{12}$ of a continuous covariate $Z$ in a model for a discrete latent class variable $\eta$. For each setting, the table shows the simulation coverage of 95% confidence intervals for $\beta_{12}$ calculated using each standard error estimate and the corresponding point estimates of $\beta_{12}$.

| Structural covariate effect | $n$ | $R_Y^2$ | One-step asympt. | Two-step | | | | | | |
|---|---|---|---|---|---|---|---|---|---|---|
| | | | | naive | asympt. | \multicolumn{5}{c}{Simulation-based, with $M$ simulation draws: $M =$} |
| | | | | | | 50 | 100 | 200 | 500 | 1000 |
| No effect | 500 | 0.8 | 95.2 | 95.2 | 95.6 | 95.2 | 95.2 | 95.2 | 95.2 | 95.2 |
| | | 0.9 | 95.9 | 96.3 | 96.3 | 96.3 | 96.3 | 96.3 | 96.3 | 96.3 |
| | 1000 | 0.8 | 92.0 | 92.8 | 92.8 | 92.8 | 92.8 | 92.8 | 92.8 | 92.8 |
| | | 0.9 | 96.0 | 96.0 | 96.0 | 96.0 | 96.0 | 96.0 | 96.0 | 96.0 |
| | 2000 | 0.8 | 94.8 | 94.8 | 94.8 | 94.8 | 94.8 | 94.8 | 94.8 | 94.8 |
| | | 0.9 | 94.4 | 94.4 | 94.4 | 94.4 | 94.4 | 94.4 | 94.4 | 94.4 |
| Mild effect | 500 | 0.8 | 96.4 | 97.6 | 97.6 | 97.6 | 97.6 | 97.6 | 97.6 | 97.6 |
| | | 0.9 | 95.2 | 95.2 | 95.6 | 95.6 | 95.6 | 95.6 | 95.6 | 95.6 |
| | 1000 | 0.8 | 96.0 | 95.6 | 95.6 | 95.6 | 95.6 | 95.6 | 95.6 | 95.6 |
| | | 0.9 | 94.8 | 94.8 | 94.8 | 94.8 | 94.8 | 94.8 | 94.8 | 94.8 |
| | 2000 | 0.8 | 95.6 | 94.8 | 94.8 | 94.8 | 94.8 | 94.8 | 94.8 | 94.8 |
| | | 0.9 | 95.6 | 95.6 | 95.6 | 95.6 | 95.6 | 95.6 | 95.6 | 95.6 |
| Strong effect | 500 | 0.8 | 92.8 | 94.4 | 94.4 | 94.8 | 94.8 | 94.8 | 94.8 | 94.8 |
| | | 0.9 | 96.8 | 96.8 | 96.8 | 96.8 | 96.8 | 96.8 | 96.8 | 96.8 |
| | 1000 | 0.8 | 94.8 | 95.6 | 96.0 | 96.0 | 96.0 | 96.0 | 96.0 | 96.0 |
| | | 0.9 | 95.2 | 94.8 | 94.8 | 94.8 | 94.8 | 94.8 | 94.8 | 94.8 |
| | 2000 | 0.8 | 94.0 | 93.6 | 94.0 | 94.0 | 94.0 | 94.0 | 94.0 | 94.0 |
| | | 0.9 | 96.4 | 96.4 | 96.4 | 96.4 | 96.4 | 96.4 | 96.4 | 96.4 |

Note: Sizes of the covariate effects and the $R_Y^2$ statistic for items $Y_j$ are defined in the text.

# 6    Two real-data examples

In this section we further illustrate the proposed method with two applied examples. Both of them have been used in previous articles on two-step estimation, the example on latent trait models in Section 6.1 by Kuha & Bakk (2023) and the example on latent class models in Section 6.2 by Bakk & Kuha (2018). More information on these examples can be found in those articles.

The computations for both examples were carried out by the same procedures developed for this article. These use a combination of Mplus 6.12 software (Muthén & Muthén, 2010) for the parameter estimation and functions in R (R Core Team, 2024) to manage the process and assemble the final estimates. We have used the *MplusAutomation* package in R (Hallquist & Wiley, 2018) to control Mplus from R, and the *brew* package (Horner, 2011) to automatically edit the input files (in particular to repeatedly re-write the fixed values of measurement parameters for step 2 of two-step estimation). The code for the examples is included in supplementary materials for this paper.

## 6.1    Latent traits as response variables: Predictors of extrinsic and intrinsic work values

The first example was previously used by Kuha & Bakk (2023) to illustrate two-step estimation for latent trait models. The data come from the Dutch sample in the fifth wave of the European Values Study (EVS), conducted in 2017–20 (EVS, 2022). We use data from the $n = 1374$ respondents who had observed values for all the covariates and at least one of the measurement items. The focus of interest is on individuals' work value orientations, represented in two dimensions as intrinsic and extrinsic work values. They are conceptualised as two continuous latent variables. The structural model of interest is a bivariate linear model for these values given a set of observed explanatory variables. In the notation of Section 2, we thus have latent variables $\boldsymbol{\eta} = (\eta_1, \eta_2)'$ and explanatory variables $\mathbf{Z}_p$, and a structural model where $\boldsymbol{\eta}$ given $\mathbf{Z}_p$ is bivariate normal with $\eta_1 \sim N(\beta_{01} + \boldsymbol{\beta}'_{11}\mathbf{Z}, \sigma_1^2)$, $\eta_2 \sim N(\beta_{02} + \boldsymbol{\beta}'_{12}\mathbf{Z}, \sigma_2^2)$ and $\sigma_{12} = \text{cov}(\eta_1, \eta_2|\mathbf{Z})$.

Each of the two work value dimensions is measured in the survey by three binary items. The measurement model for each item given its latent variable is a binary logistic model, as in equation (4). The estimated measurement model fixes the directions of both the extrinsic and intrinsic latent variable so that higher values of them correspond to a respondent placing higher importance on that dimension of work values. The structural model is a bivariate linear model with the extrinsic and intrinsic values as the two response variables. As explanatory variables it includes the respondent's sex (coded as man or woman), age (in categories: 15–29, 30–49 or 50– years), whether or not they are in a registered partnership and/or live with a partner (yes/no), and the age of the youngest person in the household (0–5, 6-17, or older).

Estimated coefficients of the structural model are shown in Table 5. We show results for two-step estimates and, for comparison, one-step estimates. The point estimates indicate, for example, that being younger or having children is associated with higher importance of extrinsic rewards of a job, and and that intrinsic rewards tend to be more important for men and for younger people.

For the standard errors of these structural coefficients Table 5 shows, first, their estimates obtained from the asymptotic closed-form matrix formulas for the one-step and two-step estimates. It then shows estimated standard errors for the two-step estimates from our estimator which uses $M$ simulations to calculate the term $\hat{\mathbf{V}}_{1M}$ which accounts for the variance from step 1. We consider the values $M = 25$, 50 and 100, plus standard errors which omit $\hat{\mathbf{V}}_{1M}$ altogether (i.e. in effect use $M = 0$). Most of the standard errors are broadly similar, but there are some differences, mostly in the model for intrinsic world values (where the residual standard deviation is higher, and the coefficients are generally estimated less precisely). For two-step estimates, the standard errors which omit the step-1 uncertainty are in some cases noticeably smaller. In these most difficult cases a larger number of simulation draws (here $M = 100$) is needed to essentially match the standard errors from the asymptotic formula, while for other coefficients (including all in the model for extrinsic work values) this is already achieved with the small value of $M = 25$.

Table 5: Estimated coefficients for linear regression models for latent traits that describe extrinsic and intrinsic work values, given some explanatory variables. The table shows point estimates of the coefficients, as well as their estimated standard errors obtained from the closed-form asymptotic formulas and (for the two-step estimates) from the simulation-based approach proposed in this paper.

|  | One-step estimates | | Two-step estimates (with standard errors from closed-form formula and from simulation-based formula with $M$ step-2 simulations) | | | | | |
|---|---|---|---|---|---|---|---|---|
|  | est. | (as. s.e.) | est. | (as.) | ($M = 100$) | ($M = 50$) | ($M = 25$) | ($M = 0$) |
| *Model for extrinsic work values:* | | | | | | | | |
| Intercept | 1.404 | (0.238) | 1.385 | (0.231) | (0.233) | (0.234) | (0.235) | (0.211) |
| Man | 0.022 | (0.104) | 0.025 | (0.101) | (0.101) | (0.102) | (0.101) | (0.099) |
| Age (vs. 30–49) | | | | | | | | |
|    15–29 | 0.489 | (0.215) | 0.473 | (0.207) | (0.207) | (0.207) | (0.207) | (0.204) |
|    50– | −0.409 | (0.155) | −0.391 | (0.150) | (0.150) | (0.150) | (0.149) | (0.145) |
| Has partner | −0.022 | (0.169) | −0.019 | (0.163) | (0.163) | (0.163) | (0.163) | (0.163) |
| Age of youngest person in household (vs. older) | | | | | | | | |
|    0–5 | 0.571 | (0.201) | 0.542 | (0.194) | (0.194) | (0.194) | (0.192) | (0.185) |
|    6–17 | 0.381 | (0.156) | 0.367 | (0.150) | (0.151) | (0.151) | (0.150) | (0.147) |
| *Model for intrinsic work values:* | | | | | | | | |
| Intercept | 0.449 | (0.377) | 0.474 | (0.415) | (0.421) | (0.426) | (0.438) | (0.404) |
| Man | 0.638 | (0.177) | 0.657 | (0.194) | (0.196) | (0.199) | (0.203) | (0.189) |
| Age (vs. 30–49) | | | | | | | | |
|    15–29 | 0.749 | (0.366) | 0.782 | (0.405) | (0.411) | (0.418) | (0.432) | (0.393) |
|    50– | −0.546 | (0.261) | −0.584 | (0.293) | (0.299) | (0.307) | (0.323) | (0.278) |
| Has partner | 0.372 | (0.293) | 0.412 | (0.328) | (0.333) | (0.340) | (0.356) | (0.313) |
| Age of youngest person in household (vs. older) | | | | | | | | |
|    0–5 | −0.072 | (0.325) | −0.082 | (0.355) | (0.356) | (0.356) | (0.356) | (0.355) |
|    6–17 | 0.010 | (0.258) | 0.028 | (0.282) | (0.283) | (0.284) | (0.285) | (0.281) |

(Two-step) estimates of residual standard deviation of extrinsic values: 1.45; residual standard deviation of intrinsic values: 2.60; residual correlation of extrinsic and intrinsic values: 0.54.

## 6.2 Latent class as an explanatory variable: Psychological contract types and job insecurity

The second example was previously used to illustrate stepwise methods of analysis for latent class models, by Bakk et al. (2013) for adjusted three-step estimation and by Bakk & Kuha (2018) for two-step estimation. The data come from the Dutch and Belgian samples of the Psychological Contracts across Employment Situations project PSYCONES (2006). We want to examine the association between an individual's perceived job insecurity and their perception of their own and their employee's obligations in their current employment (the "psychological contract"). Job insecurity is measured by a single continuous observed variable for which higher values indicate higher perceived insecurity. Types of psychological contract are described by a latent class variable, and the structural model of interest will be a normal linear regression model for job insecurity given these latent classes. In the notation of Section 2, we thus have one latent class variable $\eta$ and one observed response variable $Z_o$, and the part of the structural model that we focus on is $Z_o|\eta \sim N(\beta_0 + \sum_{c=2}^{C} \beta_{1c} I(\eta = c), \sigma^2)$.

Psychological contract types are measured by eight dichotomous survey items. Four of them refer to whether or not the employer holds specified obligations (promises given) to the employee, and four refer to obligations by the employee to the employer similarly. In each group of four, two items further refer to relational and two to transactional obligations. The measurement model for each item can be specified as in equation (5), with $h_k = 2$ for each item. Following the choices of previous studies, we consider a model with $C = 4$ latent classes, labelled as classes of "Mutual High" obligations (where respondents feel that both the employer and the employee have given obligations to each other), "Under-obligation" (obligations by the employer but not by the employee), "Over-obligation" (obligations by the employee but not by the employer), and "Mutual Low" (few or no obligations by both parties). A sample of $n = 1431$ was used for the first step of two-step estimation, and 4 respondents were omitted in the second step because of missing value of the measure of job insecurity.

Estimates of the coefficients of the structural model are shown in Table 6. These are the coefficients $\beta_{12}, \beta_{13}$, and $\beta_{14}$ of the dummy variables for three of the latent classes, with the Mutual High class as the reference class, i.e. differences in expected level of job insecurity between a class and the Mutual High class. We again show results for two-step estimates and, for comparison, one-step estimates. The point estimates show, in particular, that expected levels of job insecurity are significantly higher among indviduals who perceive the psychological contract as one where the employer has not made a commitment to them (latent classes Mutual Low and Overobligation) than among those who believe that the employer has made such a commitment (latent classes Mutual Low and Underobligation).

For the standard errors of these structural coefficients Table 6 again shows their estimates obtained from the asymptotic closed-form matrix formulas for the one-step and two-step estimates. These are very similar, indicating that there is essentially no loss of efficiency from using two-step estimation. The rest of the table then shows estimated standard errors for the two-step estimates from our estimator which uses $M$ simulations to calculate the term $\hat{\mathbf{V}}_{1M}$ which accounts for the variance from step 1. We consider the values $M = 25, 50$ and 100, plus standard errors which omit $\hat{\mathbf{V}}_{1M}$ altogether (i.e. in effect uses $M = 0$). It can be seen that this is an example where almost all of the variance comes from step 2. The simulation-based $\hat{\mathbf{V}}_{1M}$ works well, in that it correctly accounts for the (small) contribution of variance that comes from the estimated measurement model in step 1. This is the case even with a smaller number of simulations, such as here with $M = 25$. Even this setting gives essentially the same estimated standard errors as the asymptotic formulas.

Table 6: Estimated coefficients for a linear regression model for perceived job insecurity given latent classes of types of psychological contract. Here the class of "Mutual High" obligation is the reference category. The table shows point estimates of the coefficients, as well as their estimated standard errors obtained from the closed-form asymptotic formulas and (for the two-step estimates) from the simulation-based approach proposed in this paper.

|  | Coefficient (and s.e.) of latent class (vs. Mutual High): | | |
|---|---|---|---|
| *Estimator* | Over- | Under- | Mutual |
| *(standard error est.)* | obligation | obligation | Low |
| 1-step | 0.548 | −0.161 | 0.482 |
| (closed-form variance formula) | (0.077) | (0.116) | (0.113) |
| 2-step | 0.508 | −0.112 | 0.452 |
| (closed-form variance formula) | (0.076) | (0.120) | (0.111) |
| (with 100 step-2 simulations) | (0.076) | (0.120) | (0.111) |
| (with 50 step-2 simulations) | (0.076) | (0.121) | (0.112) |
| (with 25 step-2 simulations) | (0.076) | (0.121) | (0.109) |
| (with 0 step-2 simulations, i.e. with step-2 variance only) | (0.072) | (0.112) | (0.106) |

Residual standard deviation (two-step estimate): $\sigma = 0.955$

# 7 Conclusions

The two-step method has been shown to be a useful general approach for estimating structural models of latent variable models. In this paper we aimed to add to the convenience of its usability by proposing a partially simulation-based method of estimating the variances and covariances of the two-step parameter estimates. This avoids the need for the evaluation of one piece of the previously used standard variance estimates which is their most inconvenient part for practical implementation. Our proposed procedure restores the two-step procedure to a pure two-step state, where its two steps are entirely separated. From the first step we now take forward both the point estimates of the measurement parameters of the model and a set of simulated values of them. In the second step those simulated values are then used as part of the variance estimation for the estimated structural parameters.

Theoretical considerations and our simulation studies show that the proposed method works essentially as well as the standard variance estimation. Its computational cost is that estimation of the model in step 2 needs to be repeated an additional $M$ times for the $M$ simulated sets of measurement parameters from step 1. This cost, however, is likely to be modest in many cases. Our simulations and examples suggest that small values of $M$ will often be sufficient for accurate variance estimation. If desirable, $M$ may be set to a small value for initial model exploration and then increased for final estimates.

The proposed method is in principle very general and flexible. First, it can be applied without further derivations or serious coding to any latent variable models for which two-step point estimation is feasible. Second, it can be used with different methods of estimation, not just the likelihood-based estimation that was our focus here. The two-step idea itself is more generally applicable (see Gourieroux & Monfort, 1995, Sec. 24.2.4); in essence, it requires only that its two steps use consistent methods of estimation. The decomposition (8) in Section 4.2 above then suggests that an estimate of the variance matrix of estimated strutural parameters $\tilde{\boldsymbol{\theta}}_2$ could be obtained in the same way irrespective of the method of estimation, i.e. as the sum of an estimate of the variance matrix of $\tilde{\boldsymbol{\theta}}_2$ given fixed estimate $\tilde{\boldsymbol{\theta}}_1$ of the measurement parameters and the variance matrix of estimates $\tilde{\boldsymbol{\theta}}_2^{(j)}$ given simulated values of $\tilde{\boldsymbol{\theta}}_1^{(j)}$ drawn from an appropriate sampling distribution of $\tilde{\boldsymbol{\theta}}_1$. For Bayesian estimation, implemented using MCMC sampling methods, this has already been proposed by Levy (2023) and Levy & McNeish (2025). In that case $\tilde{\boldsymbol{\theta}}_1^{(j)}$ are draws from the posterior distribution of $\boldsymbol{\theta}_1$ given step-1 information only, each $\tilde{\boldsymbol{\theta}}_2^{(j)}$ is one draw from the MCMC run for the posterior distribution of $\boldsymbol{\theta}_2$ given the data and fixed $\tilde{\boldsymbol{\theta}}_1^{(j)}$, and the point and variance estimates of $\boldsymbol{\theta}_2$ are the mean (say) and variance matrix of $\tilde{\boldsymbol{\theta}}_2^{(1)}, \ldots, \tilde{\boldsymbol{\theta}}_2^{(M)}$.

Finally, we note that the method proposed here is also somewhat similar in spirit to bootstrap variance estimation of $\tilde{\boldsymbol{\theta}}_2$. The main difference is that bootstrap would use simulation to resample the data rather than draw the $\tilde{\boldsymbol{\theta}}_1^{(j)}$ from a distribution. This would incur the extra computational cost of estimating the measurement model $M$ times rather than once. On the other hand, the bootstrap would avoid the assumption that the sampling distribution of the step-1 estimates is approximately multivariate normal.